\documentclass[apjl,iop]{emulateapj}
\usepackage[utf8]{inputenc}
\usepackage{natbib} % bibtex support
\usepackage{color}
\usepackage{xspace}    %package and command to prevent missing spaces after custom newcommands

\newcommand{\referee}[1]{#1}%{\textbf{#1}}

\newcommand{\cet}{49\,Ceti\xspace}
\newcommand{\hd}{HD\,131835\xspace}
\newcommand{\molec}{molecular gas\xspace}
\newcommand{\odisk}{outer disk\xspace}
\newcommand{\idisk}{inner disk\xspace}
\newcommand{\e}[1]{$\times10^{#1}$}

\defcitealias{Roberge2013}{R13}

\begin{document}

\title{First scattered-light images of the gas-rich debris disk around 49 Ceti}

\author{\'Elodie Choquet\altaffilmark{*1},Julien Milli\altaffilmark{2},	Zahed Wahhaj\altaffilmark{2},Rémi Soummer\altaffilmark{3},Aki Roberge\altaffilmark{4},Jean-Charles Augereau\altaffilmark{5},Mark Booth\altaffilmark{6,7},Olivier Absil\altaffilmark{8**},Anthony Boccaletti\altaffilmark{9},Christine H. Chen\altaffilmark{3},John H. Debes\altaffilmark{3},Carlos del Burgo\altaffilmark{10},William R.F. Dent\altaffilmark{11},Steve Ertel\altaffilmark{12},Julien H. Girard\altaffilmark{2},Elena Gofas-Salas\altaffilmark{13},David A. Golimowski\altaffilmark{3},	Carlos A. G\'omez Gonz\'alez\altaffilmark{8},J. Brendan Hagan\altaffilmark{3}, \referee{Pascale Hibon\altaffilmark{2}}, Dean C. Hines\altaffilmark{3},Grant M. Kennedy\altaffilmark{14},Anne-Marie Lagrange\altaffilmark{5},Luca Matr\`a\altaffilmark{14},Dimitri Mawet\altaffilmark{15,1},David Mouillet\altaffilmark{5},Mamadou N'Diaye\altaffilmark{16,3},Marshall D. Perrin\altaffilmark{3},Christophe Pinte\altaffilmark{5},Laurent Pueyo\altaffilmark{3},Abhijith Rajan\altaffilmark{17},Glenn Schneider\altaffilmark{12},Schuyler Wolff\altaffilmark{18},Mark Wyatt\altaffilmark{14}}

\email{echoquet@jpl.nasa.gov}
\altaffiltext{*}{Hubble Fellow}
\altaffiltext{**}{F.R.S-FNRS Research Associate}
\altaffiltext{1}{Jet Propulsion Laboratory, California Institute of Technology, 4800 Oak Grove Drive, Pasadena, CA 91109, USA}
\altaffiltext{2}{European Southern Observatory, Alonso de C\`ordova 3107, Vitacura, Casilla 19001, Santiago, Chile}
\altaffiltext{3}{Space Telescope Science Institute, 3700 San Martin Drive, Baltimore, MD 21218, USA}
\altaffiltext{4}{Exoplanets \& Stellar Astrophysics Laboratory, NASA Goddard Space Flight Center, Code 667, Greenbelt, MD 20771, USA}
\altaffiltext{5}{Univ. Grenoble Alpes, CNRS, IPAG, F-38000 Grenoble, France }%Univ. Grenoble Alpes, Institut de Plan\'etologie et d’Astrophysique de Grenoble (IPAG, UMR 5274), 38000 Grenoble, France}
\altaffiltext{6}{Astrophysikalisches Institut und Universit\"atssternwarte, Friedrich-Schiller-Universit\"at Jena, Schillerg\"a\ss{}chen 2-3, 07745 Jena, Germany}
\altaffiltext{7}{Instituto de Astrof\'isica, Pontificia Universidad Cat\'olica de Chile, Vicu\~na Mackenna 4860, Santiago, Chile}
\altaffiltext{8}{Space sciences, Technologies and Astrophysics Research (STAR) Institute, Universit\'e de Li\`ege, 19 All\'ee du Six Ao\^ut, B-4000 Li\`ege, Belgium}

\altaffiltext{9}{LESIA, Observatoire de Paris, PSL Research University, CNRS, Sorbonne Universités, UPMC Univ. Paris 06, Univ. Paris Diderot, Sorbonne Paris Cité, 5 place Jules Janssen, 92195 Meudon, France}
\altaffiltext{10}{Instituto Nacional de Astrof\'isica, \'Optica y Electr\'onica, Luis Enrique Erro 1, Sta. Ma. Tonantzintla, Puebla, Mexico}
\altaffiltext{11}{Atacama Large Millimeter\/submillimeter Array (ALMA) Santiago Central Offices, Alonso de C\`ordova 3107, Vitacura, Casilla 763 0355, Santiago, Chile}
\altaffiltext{12}{Steward Observatory, Department of Astronomy, University of Arizona, 933 N. Cherry Avenue, Tucson, AZ 85721, USA}
\altaffiltext{13}{ONERA, 29 Avenue de la Division Leclerc, 92320 Chatillon, Paris, France}

\altaffiltext{14}{Institute of Astronomy, University of Cambridge, Madingley Road, Cambridge CB3 0HA, UK}
\altaffiltext{15}{Department of Astronomy, California Institute of Technology, 1200 E. California Blvd, MC 249-17, Pasadena, CA 91125 USA}
\altaffiltext{16}{Laboratoire Lagrange, Universit\'e C\^ote d’Azur, Observatoire de la C\^ote d’Azur, CNRS, Parc Valrose, B\^at. H. Fizeau, 06108 Nice, France}
\altaffiltext{17}{Arizona State University, Phoenix, AZ 85004, USA}
\altaffiltext{18}{Johns Hopkins University, 3400 North Charles Street, Baltimore, MD 21218, USA}

\shorttitle{First scattered-light images of 49 Ceti's debris disk}
\shortauthors{Choquet et al.}

\begin{abstract}
We present the first scattered-light images of the debris disk around \cet, a $\sim$40~Myr A1 main sequence star at 59~pc, famous for hosting two massive dust belts as well as large quantities of atomic and \molec. \referee{The outer disk is revealed in reprocessed archival \emph{Hubble Space Telescope} NICMOS F110W images, as well as new coronagraphic H band images from the \emph{Very Large Telescope} SPHERE instrument}.  The disk extends from 1\farcs1 (65~AU) to 4\farcs6 (250~AU), and is seen at an inclination of 73\degr{}, which refines previous measurements at  lower angular resolution. We also report no companion detection larger than 3~$M_{Jup}$ at projected separations beyond 20~AU from the star (0\farcs34). Comparison between the F110W and H-band images is consistent with a grey color of \cet's dust, indicating grains larger than $\gtrsim$2~$\mu$m. Our photometric measurements indicate a scattering efficiency / infrared excess ratio of 0.2--0.4, relatively low compared to other characterized debris disks. We \referee{find} that \cet presents morphological and scattering properties  very similar to the gas-rich \hd system. From our constraint on the disk inclination we find that the atomic gas previously detected in absorption must extend to the inner disk, and that the latter must be depleted of CO gas. Building on previous studies, we propose a schematic view of the system describing the dust and gas structure around \cet and hypothetic scenarios for the gas nature and origin.
\end{abstract}


%%%%%%%%%%%%%%%%%%%%%%%%%%%%%%%%%%%%%%%%%%%%%%%%%%%%%%%%%

\section{Introduction}

During their protoplanetary phase, young circumstellar environments are composed of an optically thick disk of primordial gas and dust from an original molecular cloud. Molecular gas then dominates the system over dust. Within 10~Myr though, the gas-to-dust ratio inverts as most of the gas dissipates under the successive effects of viscous accretion onto the star, photoevaporation by stellar and interstellar radiation, and accretion by giant planets. The dust grains coagulate, grow from submicron to millimeter size, and settle in the mid-plane. \referee{Then}, the disk becomes optically thin \citep{Williams2011}. By then the system has turned into a gas-poor disk of debris, with large planetesimals on colliding orbits, producing second-generation dust particles through a destructive grinding process \citep{Wyatt2008}. 

The optically thin environment in debris disks is supposed to be hostile to the persistence of primordial \molec, which is photodissociated on very short timescales by UV radiation. Yet, a handful of young systems exhibit a substantial amount of CO gas while having ages and  dust properties of debris disks \citep{Zuckerman1995,Moor2011,Dent2014,Moor2015b,Lieman-Sifry2016,Greaves2016,Marino2016}.
They all harbor large quantities of dust,  indicated by their fractional infrared luminosity $L_{IR}/L_\star \sim$1\e{-3} and have ages between 15 and 50~Myr  \citep{Greaves2016}. The nature and origin of their \molec are not completely clear: it may be primordial gas preserved from photodissociation by self-shielding, or second-generation gas released by colliding comets or planetesimals. \referee{These are} fundamental questions to investigate, as gas \referee{plays} a major role in the planet formation process.

\cet is a A1V star at 59$\pm$1~pc \citep{vanLeeuwen2007} associated with the $\sim$40~Myr Argus association \citep{Zuckerman2012}. \referee{This gas-rich system, long known for its large infrared excess \citep{Sadakane1986, Jura1998}, is now well characterized by two dust populations  \citep[][hereafter R13; \citealt{Kennedy2014,Chen2014,Moor2015a}]{Roberge2013}: a warm \idisk ($\sim$155~K, $L_{IR}/L_\star \sim$2\e{-4}) and a cold \odisk ($\sim$60~K, $L_{IR}/L_\star \sim$9\e{-4})}. The thermal emission of both dust belts has been resolved, at 12.5 and 17.9~$\mu$m for the \idisk \citep{Wahhaj2007} and at 70, 100, 160, and 450~$\mu$m for the \odisk \citep[\citetalias{Roberge2013}; ][]{Moor2015a,Greaves2016}. Dust is detected up to 60~AU in the \idisk and likely depleted below 30~AU. The \odisk is detected up to 400~AU with poor constraints on its inner radius due to the poor angular resolution achieved in the far-infrared. All images show \referee{substantially} inclined disks, albeit with poor constraints on the inclination (45--85\degr).

Both molecular and atomic gas were detected in the system. CO emission was resolved in the \odisk at similar radial scales as the dust and appears edge-on \citep{Hughes2008}. It was not detected in the \idisk, where photochemistry models confirm that \molec should be photodissociated. Despite the substantial quantity of CO ($\gtrsim$2.2\e{-4}~$M_\earth$), no CO absorption was detected in the UV \citep{Roberge2014}, indicating that the molecular disk might not be quite edge-on. Atomic gas was also detected, in the form of \ion{C}{2} emission, as well as many absorption lines (\ion{Ca}{2}, \ion{O}{1}, \ion{C}{1}, \ion{C}{2}, \ion{C}{4}, \ion{Fe}{2}), \referee{some of which show significant variability, indicative of} infalling star-grazing comets \citep{Montgomery2012,Roberge2013,Roberge2014,Malamut2014,Miles2016}. 

Analyzing the dust composition may cast a new light on the processes at work in \cet. As suggested by \citetalias{Roberge2013}, the high C/O ratio might be explained by photodesorption or grain-grain collision of carbon-rich dust. Although scattered-light images could help analyze its dust properties, \cet's disk has so far eluded near-infrared imagers, hinting \referee{at} dust grains with very low scattering efficiency. 

We present the first scattered light-images of \cet's \odisk. The disk was imaged by reprocessing archival \emph{Hubble Space Telescope} (HST) NICMOS data in the F110W filter, then with the \emph{Very Large Telescope} (VLT) SPHERE instrument in H band. The images provide a first J-H color measurement for \cet dust grains. This detection adds to the nine debris disks previously reported from the ALICE program \citep{Soummer2014,Choquet2016}, and to the detection of the debris disk around HD\,114082 from the SHARDDS program (Wahhaj et al. 2016, in press). We also report exoplanet detection limits obtained with SPHERE.

%%%%%%%%%%%%%%%
\section{Observations}

The HST image comes from an archival NICMOS dataset obtained on UT-2004-12-30 as part of a survey looking for circumstellar disks  (HST-GO-10177, PI G. Schneider). Owing to a previous non-detection \citep{Weinberger1999}, \cet was used as a reference star for PSF subtraction for the A- and B-type targets of the  survey. The images were acquired with the NIC2 camera (0\farcs07565 pixel$^{-1}$) in the F110W filter, using the 0\farcs3-radius coronagraphic mask. As it was selected as a reference star, \cet was observed at only one orientation of the spacecraft. The total exposure time is 2336~s (11 exposures).

The first SPHERE \citep{Beuzit2008} dataset was obtained on UT-2015-10-03 as part of the \emph{SPHERE High Angular Resolution Debris Disk Survey} (SHARDDS) program, a search for circumstellar disks around nearby stars with large infrared excesses (VLT 096.C-0388(A) and 097.C-0394(A), PI J. Milli). 
The images were acquired with the IRDIS instrument \citep[][0\farcs01225  pixel$^{-1}$]{Dohlen2008}  with an apodized Lyot coronagraph of radius 0\farcs0925. The target was observed in pupil-stabilized mode, using the H broad-band filter.  The total exposure time is 1920~s (480 exposures) and the field rotation through the observations was 53\degr.

Additional SPHERE data were acquired on UT-2015-10-04 in IRDIFS mode to search for giant planets within the disk (VLT 096.C-0414(A), PI M. Booth). 
Exposures were obtained with IRDIS in dual-band mode \citep{Vigan2010} with two narrow-band methane filters (H2: 1.593~$\mu$m, H3: 1.667~$\mu$m), while low resolution spectra were simultaneously obtained across the Y-J bands with the IFS \citep[39 spectral channels between 0.95 and 1.38~$\mu$m;][]{Claudi2008}. The total sky rotation achieved  in pupil-tracking mode was 105.4\degr{} in 416 exposures (6656~s total exposure time).

Satellite spots were imprinted on the first and last images of each SPHERE sequence by its deformable mirror to locate the star and field rotation center. Unocculted stellar PSFs were acquired to calibrate the instrument photometric responses.

\begin{figure*}
\center
\includegraphics[height=22cm]{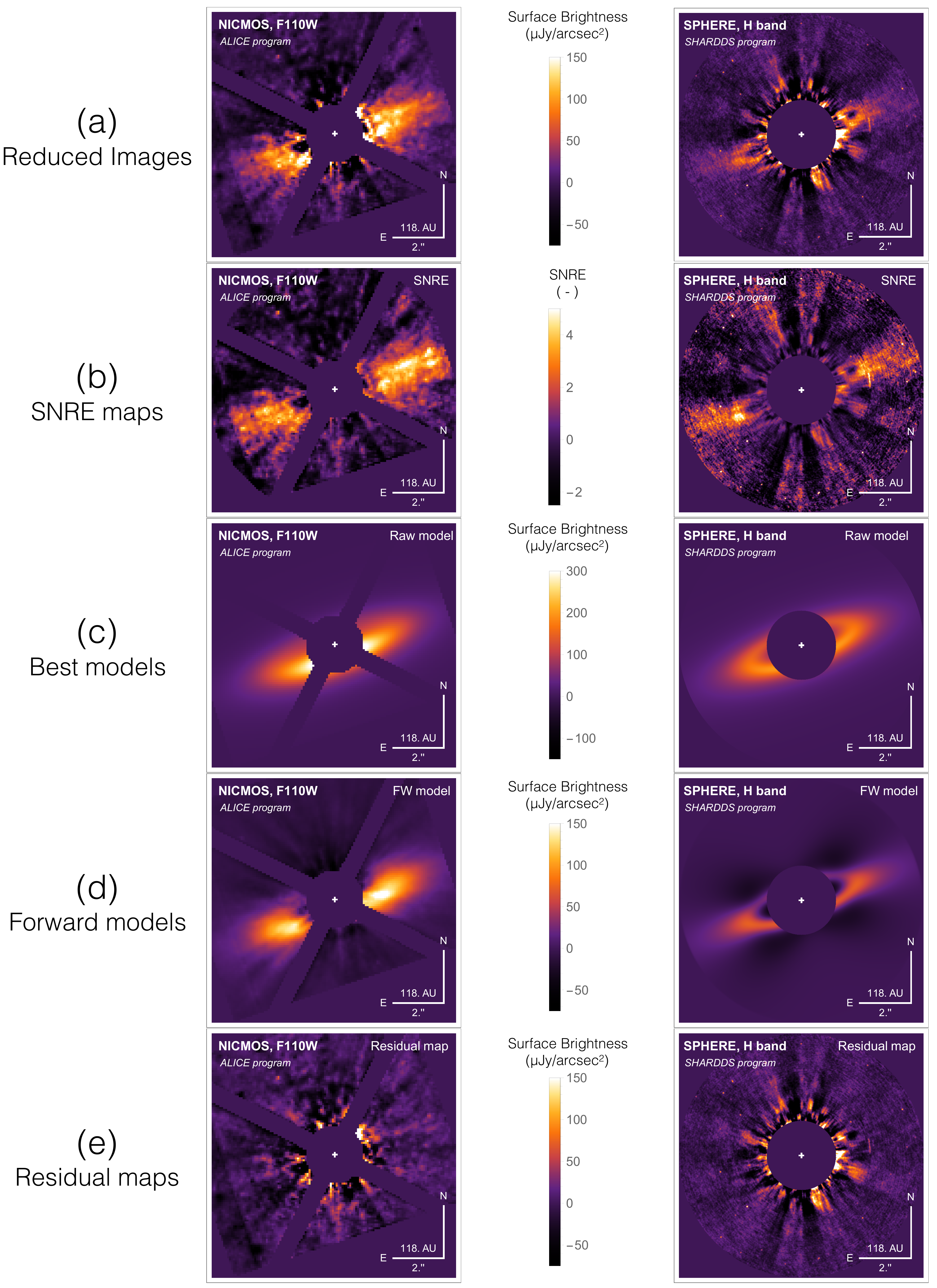}
\caption{Scattered-light images of \cet's \odisk obtained with HST-NICMOS in the F110W filter and VLT-SPHERE in H band. (a) Reduced images of the disk, smoothed by convolution with an unocculted PSF. (b) Signal-to-noise ratio per resolution element maps. (c) Best models of the disk derived from forward modeling. (d) Best models after forward modeling, smoothed by convolution with a PSF. (e) Residual maps obtained by subtracting the best forward models (d) from the reduced images (a).
 \label{fig:images}
}
\end{figure*}

%%%%%%%%%%%%%%
\section{Data Reduction}

The NICMOS data were processed as part of the \emph{Archival Legacy Investigations of Circumstellar Environments} ALICE program (HST-AR-12652, PI R. Soummer), and reduced with the ALICE pipeline \citep{Choquet2014d}. We assembled a library of 740 images, gathering the F110W PSFs of all the reference stars contemporaneous to our dataset in the NICMOS archive (58 stars). Using the 80\% of this library  most correlated with \cet data, we used the KLIP algorithm \citep{Soummer2012} to subtract the star PSFs from the images, built out of the 136 first eigenmodes of the library (23\%). The area within a radius of 10 pixels from the star was excluded from the reduction. The final image results from the mean of the individual exposures weighted by their exposure times. The signal-to-noise ratio per resolution element (SNRE) is computed as described in \citet{Rodigas2012}. 

Each SPHERE dataset was preprocessed with the SPHERE Data Reduction and Handling (DRH) pipeline \citep{Pavlov2008}.  
To image and characterize the disk, the IRDIS H-band data were reduced by subtracting the mean of the data-cube from each individual image \citep[cADI,][]{Marois2006}. No exclusion angle was used to limit disk self-subtraction effects, as the processing throughput can be recovered with forward-modeling. The final image results from the mean-combination of all derotated reduced-images. 
The SNRE map was also computed as in \citet{Rodigas2012}. The noise estimations in both NICMOS and SPHERE datasets are thus directly comparable. 
The reduced images and the SNRE maps are shown in Fig.~\ref{fig:images} (a, b).

To detect planet candidates, each SPHERE dataset was filtered to removed circularly symmetric features like the stellar halo as described in \citet{Wahhaj2013}. For speckle-subtraction, the best 60 images matching the speckle pattern of each science image between 0\farcs2 and 0\farcs6 were selected as reference images (in the speckle-aligned frames for the IFS). No more than 25\% of them were allowed to overlap  with the science image to within 1$\lambda/D$, to mitigate self-subtraction of astrophysical signal (Wahhaj et al.\ 2016 in press). The median of these reference libraries was subtracted  from the corresponding science image, then all reduced images were derotated and median-combined. Only the H2 data were combined out of the H2-H3 dataset, as the H3 frames  only contribute to the noise for methane-rich planets (although the H3 frame were included in the reference libraries to improve speckle subtraction). 

%%%%%%%%%%%%%%

\section{Results}
\cet's disk is detected \referee{at high confidence level in both datasets, with signal-to-noise ratios of 56 and 156 integrated over the whole disk respectively in the NICMOS and SPHERE images.} The SPHERE image has an angular resolution 2.3 times better than the NICMOS one. \referee{The disk is detected between 1\farcs7 and 4\farcs6 (100--275~AU) from the star, reaching SNRE$\ge5$ in some resolution elements}. This corresponds to the same extent detected in thermal emission, albeit with a much finer angular resolution. The brightest parts of the disk (SNRE $\ge$4) directly overlap with the CO emission resolved by \citet{Hughes2008}. The disk is seen at an inclination significantly different than edge-on, consistent with the estimation by \citet{Moor2015a}. We do not detect the \idisk, located at separations dominated by starlight residuals behind the reduction masks ($\le1$\arcsec).

\begin{figure}
\center
\includegraphics[width=8.5cm]{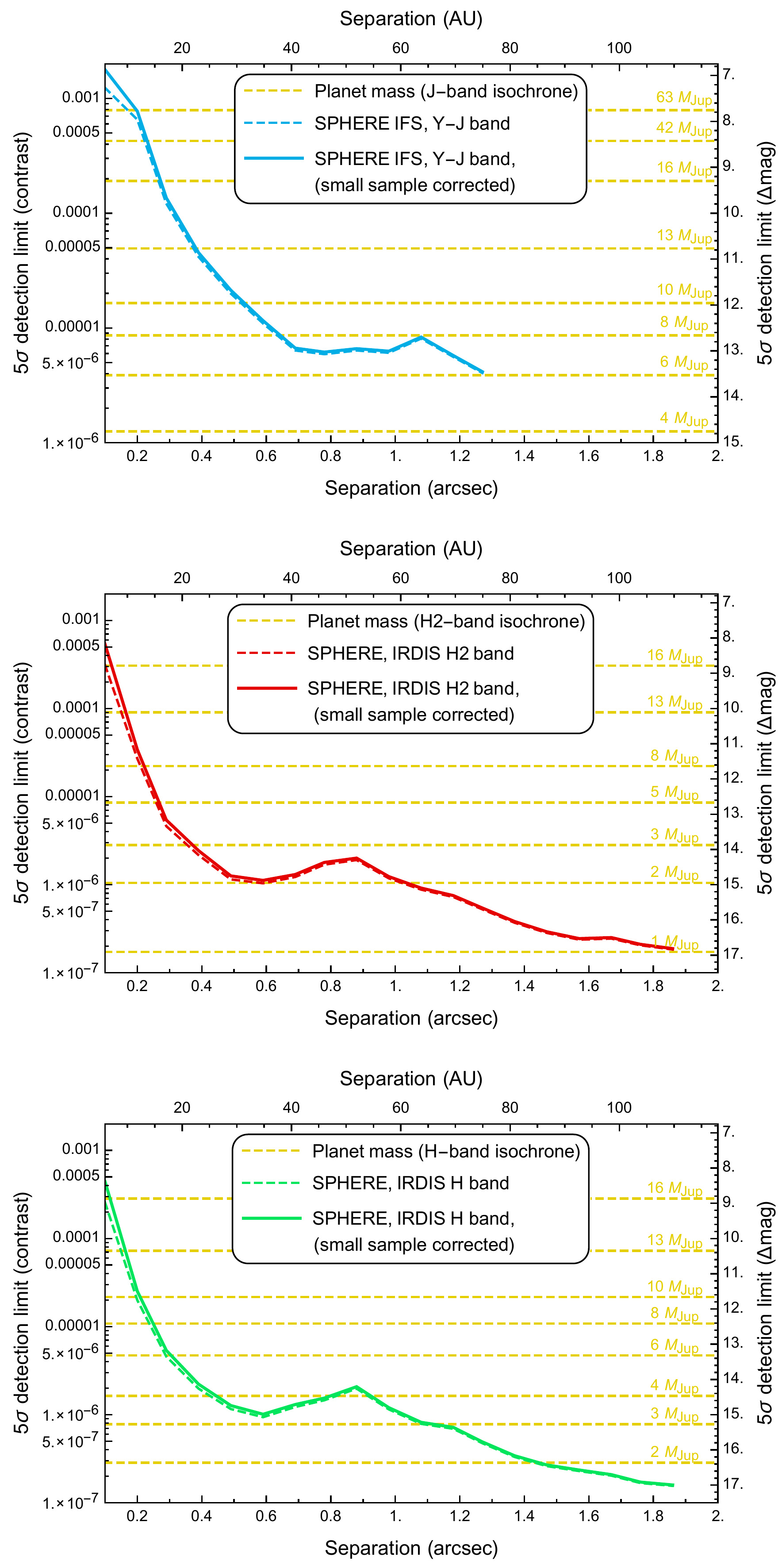}
\caption{Point source detection limits around \cet inferred from the SPHERE data \referee{corrected from small-sample statistics \citep{Mawet2014}}. Planet mass conversions are from Cond evolutionary models.\label{fig:contrast}
}
\end{figure}

No point source is detected in the datasets. Fake sources were injected in the SPHERE raw data to estimate our detection limits (Fig.~\ref{fig:contrast}). The sources had contrasts uniform across Y-J (worst case scenario), and were 10 times brighter in H2 than in H3 to simulate methane-rich planets (best case scenario). Using AMES-Cond evolutionary models \citep{Baraffe2003}, we rule out companions more massive than $3~M_{Jup}$ beyond 20~AU (within and beyond the \idisk), and than $1~M_{Jup}$ beyond 110~AU (within the \odisk).

\section{Analysis} 
\subsection{Disk modeling}\label{sec:modeling}

We analyzed our images with a model grid fitting procedure. Our objectives are: 1) constrain the main morphological characteristics of the disk; 2) investigate the significance of a possible East-West brightness asymmetry and of a continuous versus ring-like radial profile with respect to PSF-subtraction artifacts; 3) quantify the disk surface brightness and color. We modeled the NICMOS and the SPHERE images independently, using simple morphological disk models. The dust properties will be analyzed in a subsequent publication.

We used the GRaTer radiative transfer code \citep{Augereau1999a,Lebreton2012} to create scattered light images of optically thin centro-symmetric disks assuming \citet[hereafter HG]{Henyey1941} anisotropic scattering. We assumed a Gaussian profile for the vertical dust density distribution with a constant aspect ratio $h=0.05$ \citep[expected for unperturbed debris disks,][]{Thebault2009}, and we varied 6 parameters to fit the NICMOS image: the parent belt radius $R_0$, the inclination $i$, position angle (PA) $\theta$, the HG factor of scattering asymmetry $|g|$, and the radial density distribution power laws $\alpha_{in}$ and $\alpha_{out}$, respectively inward and outward from the parent radius. We varied the same parameters to fit the SPHERE image, except the PA that we fixed to 110\degr, the best value found for the NICMOS image, to decrease the computation time. The boundaries and sampling of each grid are described in Table~\ref{tab:grids}.

To calibrate the PSF-subtraction throughput and artifacts, we used the  forward modeling methods described in \citet{Choquet2016} and  \citet{Milli2012} for the NICMOS and SPHERE images respectively. The forward model intensities were scaled to the same fluxes as the images within ellipses slightly larger than and including the disk, and reduced chi square values $\chi^2_{red}$ were computed within the same area: 1\farcs8$\times$5\farcs2 semi-minor and -major axes in the NICMOS image, and 1\farcs4$\times$5\farcs4 in the SPHERE image. Circular areas of radii 1\farcs1 and 1\farcs4 from the star, respectively, were excluded.
The corresponding numbers of degrees of freedom ($N_{dof}$) are respectively 3571 and 103790. 
The best models, forward models, and residual maps are shown in Fig.~\ref{fig:images} (c, d, e). 
The parameters of the best models of the grids are reported in Table~\ref{tab:grids}, along with refined values and uncertainties computed by interpolating the $\chi^2_{red}$ values around the best model. The uncertainties correspond to the $1\sigma$ deviation expected for a chi-square distribution ($\sqrt{2N_{dof}}$).

The fits are globally consistent with each other, although the SPHERE image fit presents a smaller disk radius ($129^{+10}_{-9}$) than the NICMOS image fit ($166^{+17}_{-15}$). This discrepancy may come from degeneracies with the surface density power law parameters, as our uncertainties do not account for correlations between parameters. Both values are globally consistent with the dust thermal emission and CO gas emission extents. 
The SPHERE image fit prefers more isotropic scattering ($|g|=0.11\pm0.06$) than the NICMOS fit ($|g|=0.27\pm0.10$). This difference may be related to the dust properties.  
Both disk images are well fit by axisymmetric disk models, which demonstrates that the East/West brightness asymmetry observed in the NICMOS image is an over-subtraction artifact. Our best models show that the disk is inclined $73\pm3$\degr{} from face-on with a PA of $109\pm2$\degr{}. These values are consistent with published geometries of both the inner and outer dust belts, but significantly differ from the 90$\pm$5\degr{} inclination found for the CO gas emission by \citet{Hughes2008}. 
The NICMOS image fit favors a shallow inward slope for the disk ($\alpha_{in}=1.0^{+0.7}_{-0.9}$). The best SPHERE image fit has a consistent value, although more poorly constrained due to ADI post-processing artifacts. It is thus still unclear wether the gap seen in the SPHERE image is real or not.

\begin{deluxetable}{lccl|cc}
\tabletypesize{\scriptsize}
\tablecaption{\cet's disk modeling \label{tab:grids}}
\tablehead{
 \colhead{Param.}	& \colhead{Min.}	&\colhead{Max.}	& \colhead{$N_{val}$}			& \colhead{Best Model} & \colhead{Best Model}\\
  \colhead{}	& \colhead{}	&\colhead{}	& \colhead{}			& \colhead{} & \colhead{(interpolated)}
}
\startdata
\hline 
\multicolumn{6}{c}{NICMOS image modeling: 7680 models, $N_{dof}=3571$}\\
\hline
$R_0$ (AU)           		& 150 	&190	    	&5			&170		&$166^{+17}_{-15}$\\%160, 170,180\\
$i$ (\degr)				&68		&78		&6			&74		&$74.5^{+2.6}_{-3.2}$\\%72, 74, 76\\
$\theta$ (\degr)			&106		&112		&4	   		&110		&$109\pm2$\\%108, 110\\
$|g|$                     		&0.1       	&0.4     	&4  	     		&0.3 		&$0.27\pm0.10$\\%0.2, 0.3\\
$\alpha_{in}$ 			&0	        &3		&4			& 1   		&$1.0^{+0.7}_{-0.9}$\\%1, 2 	\\
$\alpha_{out}$ 			&-5	        &-2		&4			&-3		&$-3.2^{+0.7}_{-0.8}$\\%-3\\
\hline
$\chi^2_{red}$		&-&-&-					   	 &0.923	&-\\

\hline 
\multicolumn{6}{c}{SPHERE image modeling: 4032 models, $N_{dof}=103790$}\\
\hline
$R_0$ (AU)           		& 120 	&180	    	&7			& 130	&$129^{+10}_{-9}$\\%140\\
$i$ (\degr)				&70		&80		&6			&72		&$73\pm3$\\% 76, 78\\
$\theta$ (\degr)			&110		&110		&1	     		&110	 	&-\\
$|g|$                     		&0.0       	&0.3     	&4  	   		&0.1  	&$0.11\pm0.06$\\%0.1, 0.2\\
$\alpha_{in}$ 			&1	        &6		&6			& 3   	 	&$2.6^{+2.1}_{-1.3}$\\%3, 5, 7\\
$\alpha_{out}$ 			&-4	        &-1		&4			&-2	 	&$-2.1\pm0.5$\\%-3, -2\\
\hline
$\chi^2_{red}$		&-&-&-					   	 &0.966	&-
\enddata
\end{deluxetable}

\subsection{Disk photometry}

Using the best model images (Fig.~\ref{fig:images}c), we can estimate the disk photometry without being affected by PSF-subtraction biases. We measured disk integrated fluxes $F_{scat}$ of $2.5\pm0.9$~mJy in NICMOS F110W filter, and $2.0\pm0.7$~mJy in the SPHERE H-band filter. Given the star luminosity in these bandpasses (11.9~Jy and and 6.4~Jy respectively), we estimate that the disk has a scattering efficiency of $F_{scat}/F_{star}\sim(2.1\pm0.7)$\e{-4} in F110W and $\sim(3.2\pm1.2)$\e{-4} in H band. 
As the \odisk fractional infrared luminosity is estimated to $L_{IR}/L_{star}\sim$9.0\e{-4} \citep{Moor2015a}, we computed a ratio $f_{scat/IR}$ of the integrated scattering efficiency over the integrated infrared excess of the disk of 0.2--0.4. This ratio provides a degenerate combination of the dust albedo and phase function and is independent of the disk optical depth. 
Compared to other disks  \citep[$f_{scat/IR}\sim$0.75 on average in the visible,][Fig.~8]{Schneider2014}, \cet has a relatively low value which may point to dust with a low albedo and/or isotropic scattering.

\begin{figure}
\center
\includegraphics[width=9cm]{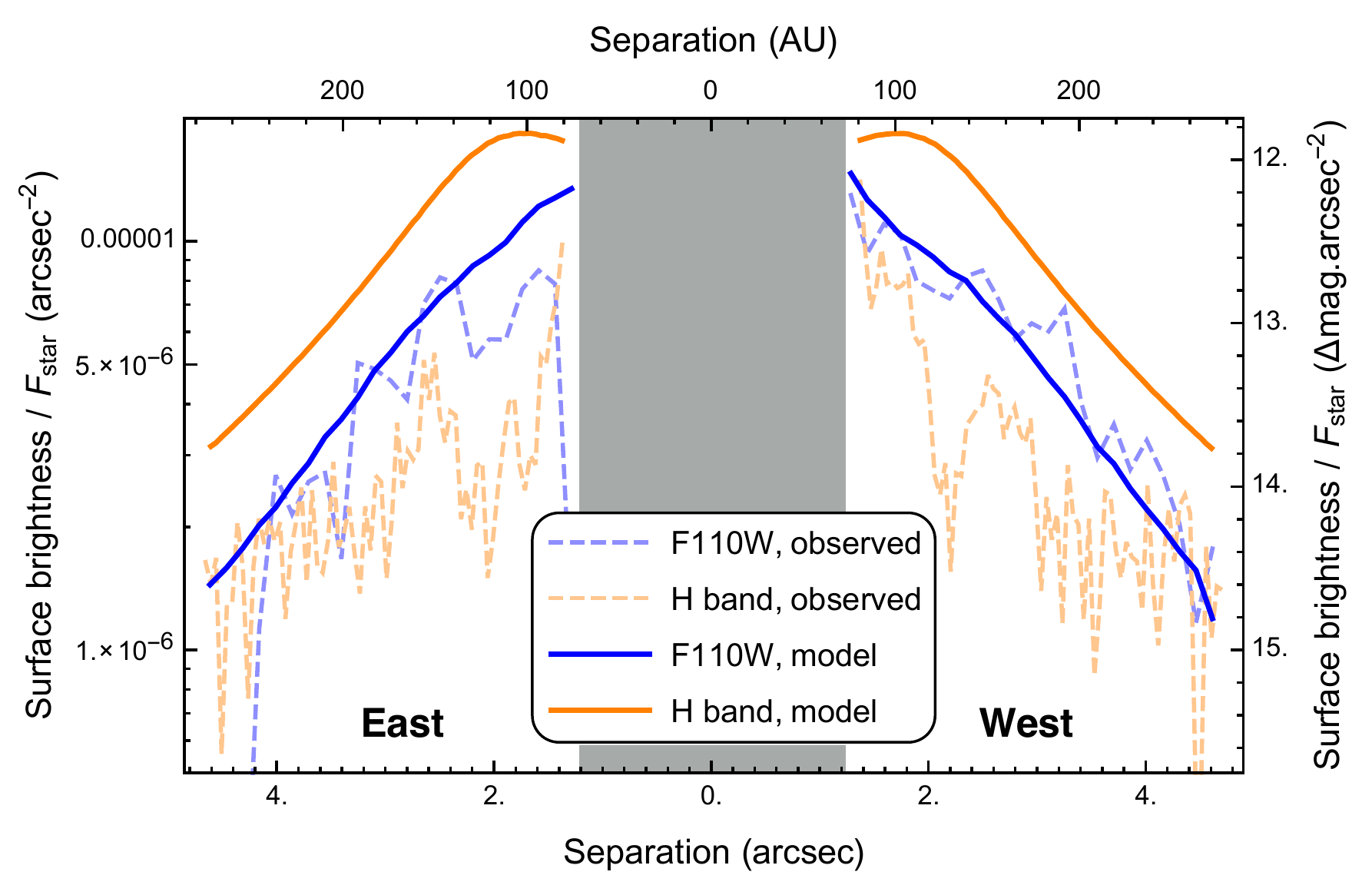}
\caption{Disk reflectance radial profiles of \cet averaging forward and backward scattering, in NICMOS-F110W filter (blue) and SPHERE H-band filter (orange). The dashed and solid lines show the reflectance measured in the reduced images and in the best models, respectively. The increasing inward profile comes from the anisotropic scattering and morphology of the disk. \label{fig:profiles}
}
\end{figure}

We also estimated the surface brightness (SB) of the disk in the two best models. In its brightest areas, (within a factor 5 of the peak values, 7.1~arcsec$^2$  and 14.2~arcsec$^2$ respectively), 
the disk's average SB is $129\pm14$~$\mu$Jy.arcsec$^{-2}$ in F110W, and $100\pm35$~$\mu$Jy.arcsec$^{-2}$ in  H band. In Fig.~\ref{fig:profiles} we present the disk SB radial profile normalized by the stellar flux. The SB profile was computed in rectangular apertures of size 1$\lambda/D \times 2\farcs3$  along the disk's major axis, averaging the forward and backward scattering along the minor axis. As for the average SB and the integrated flux, the radial profiles show a slightly red color for the dust,  consistent with a grey color.

%%%%%%%%%%%%%%

\begin{figure*}
\center
\includegraphics[width=\linewidth]{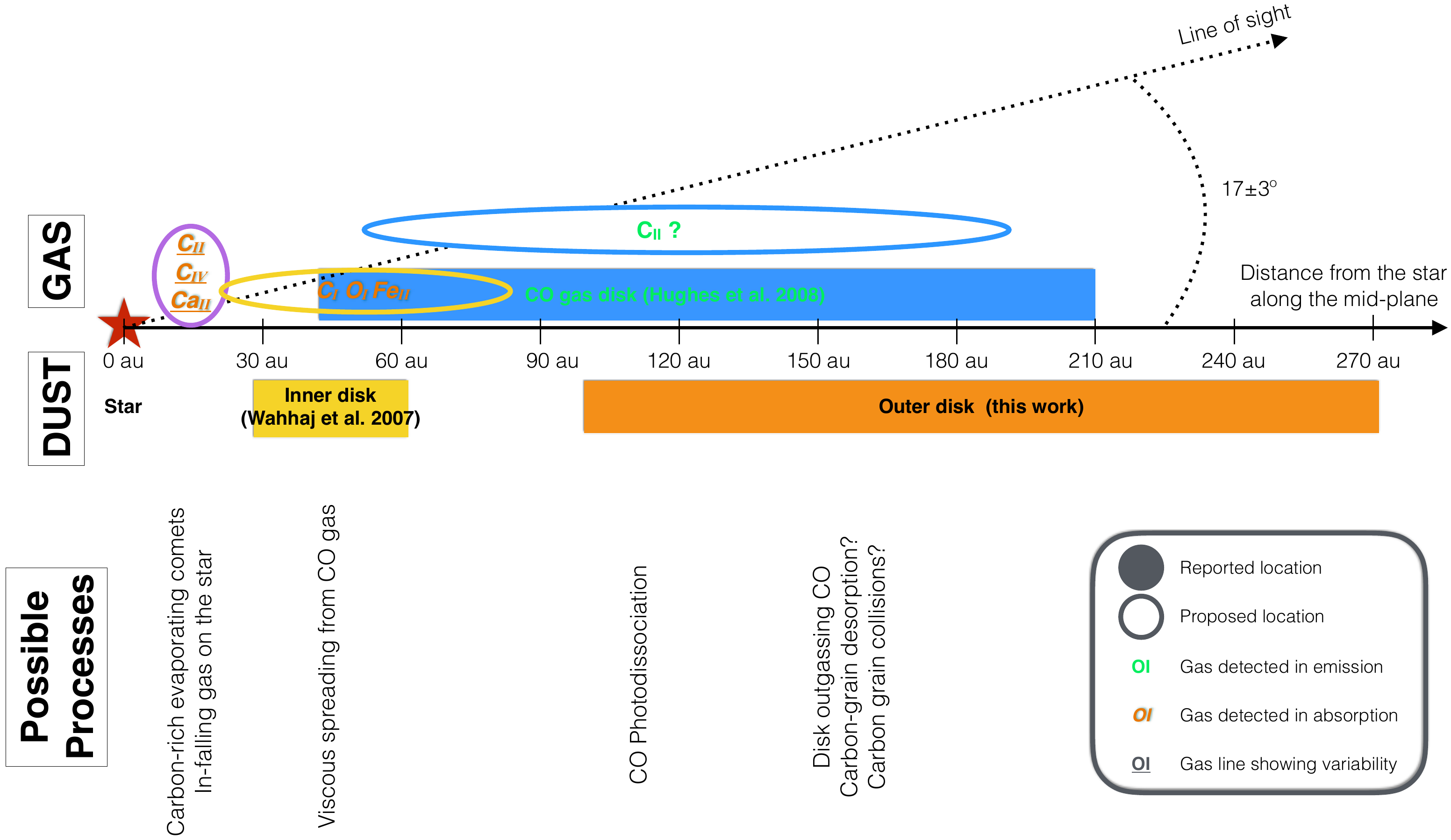}
\caption{\referee{Schematic radial view of \cet's system through the mid-plane assuming an axisymmetric disk, gathering results from this work, previous studies, and thermodynamical model scenario}. Atomic gas detections are from \citet{Montgomery2012,Roberge2013,Roberge2014,Malamut2014,Miles2016}.  \label{fig:schema}}
\end{figure*}

\section{Discussion\label{sec:discussion}}

\subsection{Similarities with \hd}

\cet's system shows interesting similarities with \hd, a 16~Myr A2IV subgiant star \citep{deZeeuw1999,Pecaut2012} also harboring a debris disk system. 
Both have high fractional infrared luminosities, well characterized by two dust populations at comparable black-body temperatures and radii ($\sim$40~AU for the \idisk, and $\sim$150~AU for the \odisk), and with comparable  dust masses $\sim$0.3~$M_{\earth}$ in the \odisk \citep{Hung2015, Moor2015a}. Both exhibit a significant amount of CO gas \citep{Moor2015b,Hughes2008}. 

Our images of \cet now reveal that the outer disks in these two systems indeed extend over the same radial distance, and are seen with the same $\sim$75\degr inclination, which enables direct comparison of their scattering properties. Both disks seem to have similar anisotropy of scattering, especially when comparing our NICMOS \cet image with the GPI \hd image, as none of them are biased by ADI artifacts. This may suggest  that the two systems have similar dust compositions. However, \hd was only detected in polarized intensity, which may bias the comparison. Complementary images of both systems would be useful to confirm these similarities, as well as a combined study of  \cet images and SED. We note that \hd's data are well fit by non-porous grains composed by a mixture of 1:1 carbon and silicate. Assuming both pure silicate and a 1:1 silicate-carbon mixture, we find with the GRaTer code that hard spheres larger than about 2~$\mu$m have a grey scattering behavior similar to what is observed for \cet. 
This grain size is  consistent with the 2.5~$\mu$m blowout size expected for silicate grains around a 2~$M_\sun$ star with luminosity $20.97~L_\sun$, and the 8~$\mu$m blowout size of the silicate-carbon mixture. 
This indicates that gas drag is likely not affecting the grain dynamics even at the tail of the collisional cascade.

\subsection{Dust and gas location}

Our images of \cet at high angular resolution unambiguously show that the \odisk is not edge-on. This result has  implications on the location of the  stable atomic (C, O, Fe) gas detected, and CO gas \emph{not} detected in absorption. 

It indicates that the atomic gas may extend close to the star to intercept the line of sight (presumably within and/or interior to the \idisk) with a higher scale height than the dust. 
%Its origin may be comets outgassing CO photodissociated in very short timescales. 
Conversely, as a significant amount of CO  is detected in emission but not in absorption, it must be confined in the mid-plane at further radii, presumably in the \odisk. This is confirmed when comparing our images with the CO renzogram from \citet{Hughes2008}.
This atomic/molecular gas segregation is consistent with the scenario of second-generation gas production proposed by \citet{Kral2016}, which can reproduce all gas observations in the $\beta$\,Pic system \citep{Matra2016}: because of very short photodissociation timescale, CO gas would only be located where it originates (outgassing comets in the \odisk), while atomic C and O gas, having much longer lifetimes, would spread inward and outward from viscous diffusion.

Our images do not clarify the location and origin of the \ion{C}{2} gas emission. It may also result from CO-photodissociation, as well as from carbon-rich dust photodesortion or grain-grain collision as suggested by \citetalias{Roberge2013}. Further analysis of \cet's dust composition would help investigate these questions. 

We propose in Fig.~\ref{fig:schema} a likely scenario for the gas and dust structure in \cet, and physical mechanisms at work in the system. It summarizes previous findings as well as results from this work.

\acknowledgments
EC acknowledges support from NASA through Hubble Fellowship grant HST-HF2-51355 awarded by STScI, operated by AURA, Inc. under contract NAS5-26555, \referee{and support from HST-AR-12652, for research carried out at the Jet Propulsion Laboratory, California Institute of Technology}.
JM acknowledges ESO through the ESO fellowship program. 
MB acknowledges support from DFG project Kr 2164/15-1. 
GMK is supported by the Royal Society as a Royal Society University Research Fellow.
CdB is supported by Mexican CONACyT research grant CB-2012-183007.
LM acknowledges support by STFC through a graduate studentship.
JCA acknowledges support by the Programme National de Planétologie.
We acknowledge support by the European Union through ERC grant 337569 for OA and CAGG, and grant 279973 for MW and LM.
%This research has made use of NASA's Astrophysics Data System, and  of the SIMBAD database, operated at CDS, Strasbourg, France.
%{\it Facilities:} \facility{HST (NICMOS), VLT (SPHERE)}.

%\bibliographystyle{apj.bst} 
%\bibliography{/Users/echoquet/Documents/Biblio/biblio-disk-planet.bib}

\end{document}